\documentstyle[11pt,newpasp,twoside,natbib,epsfig]{article}
\index{pulsars}
\index{searches}
\index{techniques}

\reversemarginpar

\begin{document}
\title{Fast Search Techniques for High Energy Pulsars}

\author{Scott M. Ransom}

\affil{McGill University Physics Dept., 3600 University St., Montreal,
  QC, H3A~2T8, Canada}

\begin{abstract}
  Modified versions of two ``standard'' pulsar search techniques are
  presented that allow large-scale searches for pulsations in long
  duration high-energy data sets using relatively modest amounts of
  computer time.  For small numbers of photons ($N_{phot} \la 10^4$),
  optimized brute-force epoch folding searches are preferred.  For
  larger numbers of photons, advanced Fourier domain acceleration
  searches are used.  Using these techniques, my collaborators and I
  have searched \emph{Chandra} observations of the Cas~A supernova
  remnant (SNR) point source and the isolated neutron star
  RX~J1856.5$-$3754 for pulsations, and confirmed the 65.6\,ms pulsar
  in the 3C~58 SNR during a blind search of archival \emph{RXTE} data.
\end{abstract}

\section{Introduction}

Sensitive searches for pulsars in long duration high-energy (i.e.
X-ray and $\gamma$-ray) observations are difficult.  For relatively
short observations ($T_{obs} \la 20$\,ks) all known isolated pulsars
can be assumed to have a constant spin period during the observation.
In this limiting case, searches for relatively bright pulsars are
usually implemented using simple epoch folding or a standard Fourier
analysis \citep[e.g.][]{lde+83,van89}.  When $T_{obs}$ is long,
however, the spin-down of young high-energy pulsars becomes important.
The corresponding change in frequency during an observation can make
these pulsars invisible to traditional search techniques.

Another difficulty involved in searching very long observations (i.e.
many days) is the extraordinarily large number of independent trials
required for a blind search.  For epoch folding searches that cover a
wide range of both frequency $f$ and frequency derivative $\dot{f}$,
the computational complexity is $\propto N_{phot}T_{obs}^2$, or
$\propto T_{obs}^3$ for a source of constant flux.  Advanced Fourier
analyses of very long time series (i.e.  $(2-20)\times 10^8$ points
where the photons have been binned) have been carried out by a few
groups \citep[e.g.][]{mkk+00, ckl+01} and are, in general, more
efficient due to the use of the FFT (computational complexity $\propto
T_{obs}^2\log T_{obs}$).  The difficulties involved in computing many
FFTs of this size, however, have relegated such analyses to those with
access to large-scale super-computing.

The techniques that I describe here allow large-scale blind searches
of very long high-energy observations to be carried out on modest
(albeit modern) workstations or inexpensive workstation clusters.

\section{Optimized Epoch Folding ($N_{phot}\la 10^4$)}

Epoch folding involves placing each event (i.e. photon) into one of a
relatively small number of pulse phase bins (usually 2$-$20) by
calculating the exact pulse phase of the event based on a trial pulsar
ephemeris (e.g. $f$ and $\dot{f}$) and an arbitrary starting phase of
the profile (i.e. phase zero).  For each trial profile, one of many
possible (but preferably computationally ``cheap'') statistical tests
is performed on the profile to determine if it shows a pulsed signal or not.

For blind searches, a large range of $f$ and $\dot{f}$ must be covered
at high resolution in both the $f$ and $\dot{f}$ directions.  In order
to maximize sensitivity to low duty-cycle (i.e. narrow) pulse
profiles, one must oversample the Independent Fourier Spacing
($1/T_{obs}$) in frequency and the equivalent independent spacing in
$\dot{f}$ ($\sim 6/T_{obs}^2$) by a factor of 4 or more.

More advanced searches also attempt to match the complexity of the
expected pulse shape with an appropriate number of pulse phase bins at
the optimal starting phase of the profile.  For example, sinusoidal
profiles only require 2 properly phased bins for good sensitivity, but
narrow pulse profiles are best detected if a single phase bin has the
same width as the pulse and is centered on it.  Unfortunately,
searching over various numbers of phase bins and starting phases
greatly increases the computer time required in an already costly
search.

I have made modifications to the Bayesian method of epoch folding
presented by \citet{gl92b,gl96} that incorporates everything described
above in an efficient manner.  These modifications allow brute force
searches of ``reasonable'' high-energy data sets in a few weeks or
less on a fast workstation (i.e.  $N_{phot} \sim 1000$\,photons,
$T_{obs} \la 30$\,days, $0 < f \la 500$\,Hz, $|\dot{f}| \la
10^{-9}$\,Hz\,s$^{-1}$).

The calculations required to search over various numbers of phase bins
and starting phases have been reduced dramatically from the exact
method described in \citet{gl92b}.  The basic idea is to initially
fold the data into 60 or 120 phase bins and then combine various
numbers of these bins (i.e.  2, 3, 4, 5, 6, 10, 12, 15, 20 bins) in an
efficient manner at each of the 60 or 120 starting phases in order to
maximize the signal-to-noise.

For very long duration observations, huge savings in CPU time can be
had by searching the photons in a hierarchical manner and accepting
some loss in sensitivity.  Since for a search of a continuously
sampled observation of a source with constant flux the required number
of operations is $\propto T_{obs}^3$, searching only a short but
continuous part of the data takes much less time than searching the
full observation (the number of photons and required resolution in $f$
and $\dot{f}$ are reduced significantly).  Candidates above a very low
threshold (say the top few percent) in this initial search are then
examined using twice as many photons (or approximately twice the
duration) at a correspondingly finer resolution.  This process
continues until all the photons are examined and the full resolution
of the data is reached around the candidate.

My collaborators and I have recently applied these techniques to
brute-force searches of \emph{Chandra} observations of two very
interesting X-ray sources: the point source found near the center of
the Cas~A SNR \citep{tan99} and the isolated neutron star
RX~J1856.5$-$3754 \citep*{wwn96}.  \citet{mrj+02} searched a 50\,ks
HRC-S observation (OBSID 1857) from the Cas~A point source and
reported no convincing candidates.  More recently, we have searched
the follow-up 50\,ks observation (OBSID 1038) of Cas~A and could
neither find any convincing new candidates nor confirm any of the low
significance candidates from the earlier search (manuscript in
preparation).  A similarly unsuccessful search for pulsations (and a
corresponding 99\% confidence level upper limit to pulsations of
4.5\%) from the 450\,ks Director's Discretionary Time observation of
RX~J1856.5$-$3754 was reported by \citet*{rgs02}.

\section{Fourier-Domain Acceleration Searches ($N_{phot}\ga 10^3$)}

When the observed number of photons is too large to allow efficient
epoch folding searches to be conducted, the photons can be binned into
a time series and searched using acceleration searches (i.e. $\dot{f}$
is constant during the observation) instead.  Time domain acceleration
searches have been used by numerous groups for some time
\citep[e.g.][]{mk84} and involve either re-sampling or re-binning the
original uncorrected data to account for a non-zero $\dot{f}$.  Each
time series corresponding to a different $\dot{f}$ trial then requires
a long FFT and a subsequent search for pulsations.  Unfortunately,
FFTs of very long time series are difficult to compute
\citep[e.g.][]{ckl+01}.

By using Fourier domain acceleration search techniques
\citep[][]{ran01}, only a single FFT of the original time series is
required.  All $\dot{f}$ trials are computed by correlating complex
template responses with the raw Fourier amplitudes from the original
FFT (i.e. matched filtering in the Fourier domain).  The single long
FFT that is required can be easily computed using out-of-core memory
FFT techniques on even modest workstations.  The subsequent Fourier
domain acceleration search is very efficient since the matched
filtering is inherently memory local and easily computed using short
FFTs.

My collaborators and I have used a Fourier domain acceleration search
code that includes Fourier interpolation \citep[to minimize the
effects of ``scalloping'', e.g.][]{van89} and harmonic summing (to
improve sensitivity to low duty-cycle pulsations) to successfully
search numerous long radio observations of globular clusters and X-ray
observations of potential X-ray pulsars.  New pulsars detected using
these methods include the binary millisecond radio pulsar
PSR~J1807$-$2459 in globular cluster NGC6544 \citep{rgh+01} and a
probably isolated 4.714\,ms pulsar in Terzan~5 \citep{ran01}.

Recently, \citet{mss+02} used these techniques to confirm the new
65.6\,ms X-ray pulsar discovered with \emph{Chandra} in the 3C~58 SNR
and measure its spin-down characteristics.  We conducted a blind
Fourier domain acceleration search on an archival 36\,ks \emph{RXTE}
observation (OBSID 20259-02-01-00 with 20.6\,ks on-source) and
uncovered a $\sim 6.4\,\sigma$ candidate as a sum of 8 harmonics (the
strongest having only $\sim 13$ times the local mean power --- showing
the need for a harmonic sum).  The pulse profile from the \emph{RXTE}
observation is shown in Figure~1.  By comparing the measured pulse
periods from the \emph{RXTE} and \emph{Chandra} data we were able to
measure the period derivative ($1.93 \times 10^{-13}$), spin-down
$\dot{E}$ ($\sim 2.7\times 10^{37}$\,ergs\,s$^{-1}$), surface magnetic
field strength ($\sim 3.6\times 10^{12}$\,gauss), and characteristic
age ($\sim 5380$\,y), showing that it is a surprisingly ``normal''
young pulsar.

\centerline{\psfig{figure=ransoms1_1.eps, height=3.25in,angle=270,clip=}}
\noindent{\footnotesize {\bf Figure 1} --- 
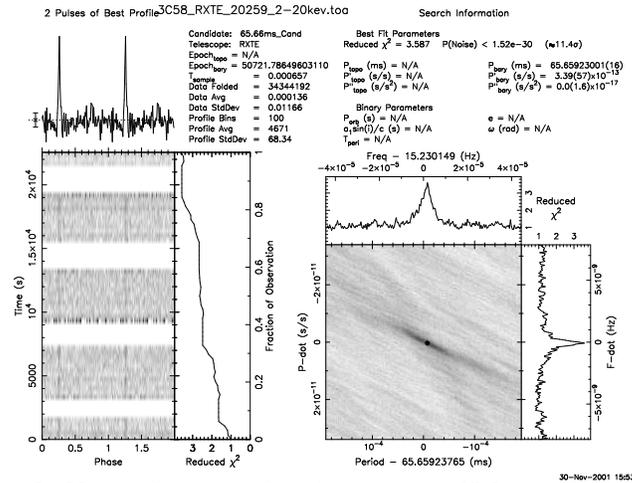
  The 2$-$20\,kev \emph{RXTE} profile of the young 65.6\,ms X-ray
  pulsar J0205+6449 recently discovered in supernova remnant 3C~58
  (see \S3).}


\begin{thebibliography}{14}

\parindent=0pt
\parskip=-1.2ex

\expandafter\ifx\csname natexlab\endcsname\relax\def\natexlab#1{#1}\fi

\bibitem[{{Chandler} {et~al.}(2001){Chandler}, {Koh}, {Lamb}, {Macomb},
  {Mattox}, {Prince}, \& {Ray}}]{ckl+01}
{Chandler}, A.~M., {Koh}, D.~T., {Lamb}, R.~C., {Macomb}, D.~J., {Mattox},
  J.~R., {Prince}, T.~A., \& {Ray}, P.~S. 2001, \apj, 556, 59

\bibitem[{Gregory \& Loredo(1992)}]{gl92b}
Gregory, P.~C. \& Loredo, T.~J. 1992, \apj, 398, 146

\bibitem[{Gregory \& Loredo(1996)}]{gl96}
---. 1996, \apj, 473, 1059

\bibitem[{Leahy {et~al.}(1983)Leahy, Darbro, Elsner, Weisskopf, Sutherland,
  Kahn, \& Grindlay}]{lde+83}
Leahy, D.~A., Darbro, W., Elsner, R.~F., Weisskopf, M.~C., Sutherland, P.~G.,
  Kahn, S., \& Grindlay, J.~E. 1983, \apj, 266, 160

\bibitem[{Middleditch \& Kristian(1984)}]{mk84}
Middleditch, J. \& Kristian, J. 1984, \apj, 279, 157

\bibitem[{{Middleditch} {et~al.}(2000){Middleditch}, {Kristian}, {Kunkel},
  {Hill}, {Watson}, {Lucinio}, {Imamura}, {Steiman-Cameron}, {Shearer},
  {Butler}, {Redfern}, \& {Danks}}]{mkk+00}
{Middleditch}, J., {Kristian}, J.~A., {Kunkel}, W.~E., {Hill}, K.~M., {Watson},
  R.~D., {Lucinio}, R., {Imamura}, J.~N., {Steiman-Cameron}, T.~Y., {Shearer},
  A., {Butler}, R., {Redfern}, M., \& {Danks}, A.~C. 2000, New Astronomy, 5,
  243

\bibitem[{{Murray} {et~al.}(2002{\natexlab{a}}){Murray}, {Ransom}, {Juda},
  {Hwang}, \& {Holt}}]{mrj+02}
{Murray}, S.~S., {Ransom}, S.~M., {Juda}, M., {Hwang}, U., \& {Holt}, S.~S.
  2002{\natexlab{a}}, \apj, in press (astro-ph/0106516)

\bibitem[{{Murray} {et~al.}(2002{\natexlab{b}}){Murray}, {Slane}, {Seward},
  {Ransom}, \& {Gaensler}}]{mss+02}
{Murray}, S.~S., {Slane}, P.~O., {Seward}, F.~D., {Ransom}, S.~M., \&
  {Gaensler}, B.~M. 2002{\natexlab{b}}, \apj, in press (astro-ph/0108489)

\bibitem[{{Ransom}(2001)}]{ran01}
{Ransom}, S.~M. 2001, PhD thesis, Harvard University

\bibitem[{{Ransom} {et~al.}(2002){Ransom}, {Gaensler}, \& {Slane}}]{rgs02}
{Ransom}, S.~M., {Gaensler}, B.~M., \& {Slane}, P.~O. 2002, \apjl, submitted
  (astro-ph/0111339)

\bibitem[{{Ransom} {et~al.}(2001){Ransom}, {Greenhill}, {Herrnstein},
  {Manchester}, {Camilo}, {Eikenberry}, \& {Lyne}}]{rgh+01}
{Ransom}, S.~M., {Greenhill}, L.~J., {Herrnstein}, J.~R., {Manchester}, R.~N.,
  {Camilo}, F., {Eikenberry}, S.~S., \& {Lyne}, A.~G. 2001, \apjl, 546, L25

\bibitem[{Tananbaum(1999)}]{tan99}
Tananbaum, H. 1999, IAU Circ., 7246

\bibitem[{{van der Klis}(1989)}]{van89}
{van der Klis}, M. 1989, in Timing Neutron Stars, ({NATO ASI Series}), ed.
  H.~\"{O}gelman \& E.~P.~J. {van den Heuvel} (Dordrecht: Kluwer), 27--69

\bibitem[{{Walter} {et~al.}(1996){Walter}, {Wolk}, \& {Neuh{\" a}user}}]{wwn96}
{Walter}, F.~M., {Wolk}, S.~J., \& {Neuh{\" a}user}, R. 1996, \nat, 379, 233

\end{thebibliography}
\end{document}